     \documentclass[12pt]{article} 
     \usepackage{epsfig}
     \newlength{\dinwidth}                       
     \newlength{\dinmargin}                      
\def\Journal#1#2#3#4{{#1} {\bf #2}, #3 (#4)}

\def\NPB{{\em Nucl. Phys.} B}
\def\PLB{{\em Phys. Lett.}  B}

\def\PRD{{\em Phys. Rev.} D}
\def\ZPC{{\em Z. Phys.} C}
\def\lsim{\mathrel{\rlap{\lower4pt\hbox{\hskip1pt$\sim$}}
    \raise1pt\hbox{$<$}}}                
\def\gsim{\mathrel{\rlap{\lower4pt\hbox{\hskip1pt$\sim$}}
    \raise1pt\hbox{$>$}}}                

\catcode`@=11

\def\@citex[#1]#2{\if@filesw\immediate\write\@auxout{\string\citation{#2}}\fi
  \def\@citea{}\@cite{\@for\@citeb:=#2\do
    {\@citea\def\@citea{,\penalty\@m}\@ifundefined
      {b@\@citeb}{{\bf ?}\@warning
       {Citation `\@citeb' on page \thepage \space undefined}}%
\hbox{\csname b@\@citeb\endcsname}}}{#1}}

\def\citer{\@ifnextchar [{\@tempswatrue\@citexr}{\@tempswafalse\@citexr[]}}

\def\@citexr[#1]#2{\if@filesw\immediate\write\@auxout{\string\citation{#2}}\fi
  \def\@citea{}\@cite{\@for\@citeb:=#2\do
    {\@citea\def\@citea{--\penalty\@m}\@ifundefined
       {b@\@citeb}{{\bf ?}\@warning
       {Citation `\@citeb' on page \thepage \space undefined}}%
\hbox{\csname b@\@citeb\endcsname}}}{#1}}
\catcode`@=12

\def\a{\alpha}
\def\b{\beta}

\def\f{\phi}
\def\g{\gamma}
\def\h{\eta}

\def\j{\psi}

\def\l{\lambda}
\def\m{\mu}
\def\n{\nu}
\def\o{\omega}
\def\p{\pi}

\def\x{\xi}
\def\z{\zeta}

\def\G{\Gamma}

\def\bo{{\raise.15ex\hbox{\large$\Box$}}}               
\def\face{{\raise.2ex\hbox{$\displaystyle \bigodot$}\mskip-2.2mu \llap {$\ddot
        \smile$}}}                                      

\def\Bar#1{\overline{#1}}                       
\def\leftrightarrowfill{$\mathsurround=0pt \mathord\leftarrow \mkern-6mu
        \cleaders\hbox{$\mkern-2mu \mathord- \mkern-2mu$}\hfill
        \mkern-6mu \mathord\rightarrow$}       
\def\dvec#1{\vbox{\ialign{##\crcr
        \leftrightarrowfill\crcr\noalign{\kern-1pt\nointerlineskip}
        $\hfil\displaystyle{#1}\hfil$\crcr}}}           

\def\beq{\begin{equation}}
\def\eeq{\end{equation}}

\def\beqx{\begin{displaymath}}
\def\eeqx{\end{displaymath}}

\def\beql{\begin{eqnarray}}
\def\eeql{\end{eqnarray}}

\begin{document}
\begin{flushright}
ANL-HEP-CP-99-35\\
NIKHEF/99-013\\
ITP-SB-99-18
\end{flushright}

\vspace*{10mm}
\begin{center}  \begin{Large} \begin{bf}
Heavy Quark Production in Deep-Inelastic Scattering at HERA\\
  \end{bf}  \end{Large}
  \vspace*{5mm}
  \begin{large}
B.W. Harris$^a$, E. Laenen$^{b}$, S. Moch$^b$, J. Smith$^c$\\
  \end{large}
\end{center}
$^a$ Argonne National Laboratory, 9700 S. Cass Ave., Argonne, IL 60439, USA\\
$^b$ NIKHEF Theory Group, Kruislaan 409, 1098 SJ Amsterdam, The Netherlands\\
$^c$ Institute for Theoretical Physics, S.U.N.Y. at Stony Brook, Stony Brook, NY 11794-3840, USA\\
\begin{quotation}
\noindent
{\bf Abstract:}
We discuss two topics in the production of heavy quarks in deep-inelastic 
scattering: the next-to-leading order Monte-Carlo {\sc HVQDIS} and 
the next-to-leading logarithmic resummation of soft gluon effects, 
including estimates of next-to-next-to-leading order corrections therefrom.
\end{quotation}
\section{Introduction}

Charm quarks produced in deep-inelastic scattering have been identified in
sizable numbers by the H1 and ZEUS collaborations at HERA \cite{HERAc}, and
considerably more charm (and bottom) data are anticipated.
At the theoretical level the reaction has already been studied extensively.
In the framework where the heavy quark is not treated as a 
parton, leading order (LO) \cite{gr} and next-to-leading order 
(NLO) \cite{lrsvn93} calculations of the inclusive structure functions
exist. Moreover, LO ({\sc AROMA, RAPGAP}) \cite{is,rapgap} and 
NLO ({\sc HVQDIS})\cite{bh,hs1} 
Monte-Carlo programs, allowing a much larger class of observables
to be compared with data, have been constructed in recent years. 
The NLO QCD description agrees quite 
well with the HERA data.  Most of the recent theoretical attention
for this reaction has been in the context of 
variable flavor number schemes \cite{vfns}, which we shall not address here.
We shall review here two issues regarding heavy quark production in 
deep-inelastic scattering.
In the next section we discuss two new features of {\sc HVQDIS}, relevant
to recent and future analyses; first, the inclusion of charmed-meson
semi-leptonic decays, and second, a switch to describe (LO)
$D$--${\overline D}$--$jet$ final states.
In the third section we review the methods and some key results
of the next-to-leading logarithmic Sudakov resummation
for DIS production of heavy quarks, and NNLO estimates derived 
from this resummation.

\section{The NLO Monte-Carlo {\sc HVQDIS}}

The program {\sc HVQDIS} \cite{bh} is based on the next-to-leading 
order fully differential heavy quark contributions to the proton 
structure functions \cite{hs1}.
The basic components (in terms of virtual-photon-proton scattering) 
are the 2--to--2 body squared matrix elements through one-loop order and 
tree level 2--to--3 body squared matrix elements, 
for both photon-gluon and photon-light-quark 
initiated subprocesses.  It is therefore possible 
to study single- and semi-inclusive production at next-to-leading 
order, and three body final states at leading order.
The goal of the next-to-leading order calculation 
is to organize the soft and collinear singularity 
cancellations without loss of information 
in terms of observables that can be predicted.

The subtraction method provides a mechanism for this cancellation.
It allows one to isolate the soft and collinear singularities 
of the 2--to--3 body processes within the framework of dimensional 
regularization without calculating all the phase space integrals in 
a space-time dimension $n\ne 4$.
Expressions for the three-body squared matrix elements in the limit 
where an emitted gluon is soft appear in a factorized form where 
poles $\epsilon=2-n/2$ multiply leading order squared matrix elements.  
These soft singularities cancel upon addition 
of the interference of the leading order diagrams with the 
renormalized one-loop virtual diagrams.
The remaining singularities are initial state collinear in origin 
wherein the three-body squared matrix elements appear in a 
factorized form, with poles in $\epsilon$ multiply splitting 
functions convolved with leading order squared matrix elements. 
These collinear singularities are removed through mass 
factorization.

The result of this calculation is an expression 
that is finite in four-dimensional space time.  One can 
compute all phase space integrations using 
standard Monte-Carlo integration techniques.
The final result is a program which returns 
parton kinematic configurations and their corresponding 
weights, accurate to ${\cal O}(\alpha\alpha_s^2)$.
The user is free to histogram any set of 
infrared-safe observables and apply 
cuts, all in a single histogramming subroutine.
Additionally, one may study heavy hadrons using the  
Peterson {\em et al}.\ model.
Detailed physics results from this program are 
given in \cite{hs2}.

Below we discuss and give examples of two new 
options of {\sc HVQDIS} version 1.3\footnote{available from 
{\tt harris@hep.anl.gov}}: electrons from semileptonic 
decays of the charmed hadron, and a switch for retaining only 
three body final states.

\subsection{Semileptonic decays}

\begin{figure}[tb]
\begin{center}
\epsfxsize= 3in  
\leavevmode
\epsfbox{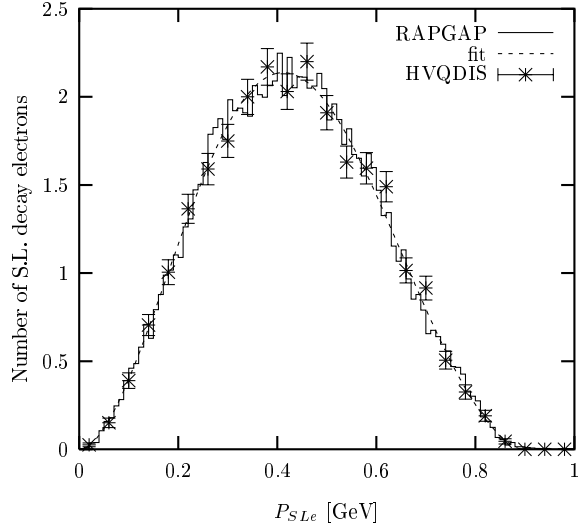}
\end{center}
\caption{Momentum distribution of semileptonic decay electrons from charmed 
hadrons produced by RAPGAP in the rest frame of the hadron.  Momenta 
distributed according to the fit are implemented in {\sc HVQDIS1.3}.}
\label{fig1}
\end{figure}

{\sc HVQDIS} has been extended to include the electron from the 
semileptonic decay of the charmed hadron.  In the rest frame of 
the decaying hadron, the electrons are distributed isotropically.  
Their momentum distribution is the weighted average of multiple 
decay modes of many different charmed hadrons, and has been 
deduced \cite{wv} from {\sc RAPGAP}\cite{rapgap}.
The parameterization is shown in fig.~\ref{fig1}.  
The implementation used in {\sc HVQDIS1.3} (shown as points) 
was derived from the fit (dashed line) to the {\sc RAPGAP} output 
(histogram).  The overall normalization of the cross section is 
then fixed by the semileptonic branching ratio 
Br$(c \rightarrow e)$ which we take to be $9.5\%$.

\begin{figure}[tb]
\begin{center}
\epsfxsize= 5in  
\leavevmode
\epsfbox{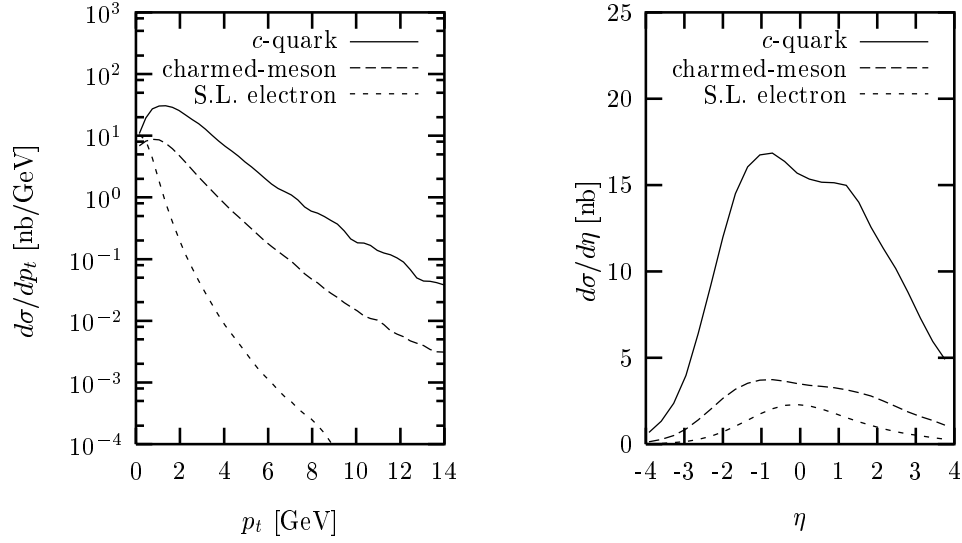}
\end{center}
\caption{The inclusive transverse momentum (left) and psuedo-rapidity (right) 
distributions of the semileptonic electron, the parent parton ($c$-quark), 
and the hadron (charmed-meson) in the lab frame for $0<y<0.7$ and $2<Q^2<100\; 
{\rm GeV}^2$.}
\label{fig2}
\end{figure}

The inclusive transverse momentum and pseudo-rapidity  
distributions of the semileptonic decay electrons in the 
lab frame in 
deep inelastic scattering of 820 GeV protons with 27.5 GeV 
electrons in the kinematic range $0<y<0.7$ and $2<Q^2<100$ GeV$^2$
are shown in fig.~\ref{fig2}.  We also show the corresponding 
distributions for the parent parton ($c$-quark) and hadron (charmed-meson).
The curves are produced using the next-to-leading order 
Gl\"{u}ck-Reya-Vogt 1994 (GRV94) \cite{grv94} parton distribution 
functions, a two-loop $\alpha_s$ with $n_f=3$ and 
$\Lambda_{\rm QCD}^{(n_f=3)}=248$ MeV, and $m_c=1.35$ GeV.
The distributions of the charmed partons and hadrons are highly 
correlated because of the simple Peterson {\em et al}. fragmentation 
model.  The semileptonic decay electrons are very soft, taking only a 
small portion of the hadron $p_t$, and more central due to the 
isotropic nature of the decay.

\subsection{Three body final states}

For speed considerations it pays to add a switch to turn off all two 
body contributions (primarily the very slow virtual routines) when 
one is interested only in a manifestly three body observable.  
Such a switch has been added to {\sc HVQDIS1.3}.  
We give here a sample of three body observables. 

The final state of interest is $D$--${\overline D}$--{\em jet} 
corresponding to the partonic states $c$--${\overline c}$--$g$ and 
$c$--${\overline c}$--$q$.  We begin by requiring the 
$D$, ${\overline D}$ and jet to be above some minimum transverse 
momentum 
($P_t^D>1.2$ GeV, $P_t^{\overline D}>1.2$ GeV, $P_t^{jet}>6$ GeV),  
to be central 
($|\eta_D|<1.5$, $|\eta_{\overline D}|<1.5$, $|\eta_{jet}|<2.4$), 
and to be isolated 
($R_{D{\overline D}}>0.7$, $R_{Djet}>0.7$, $R_{{\overline D}jet}>0.7$) 
in the lab frame.
The cone size $R_{ij}=\sqrt{(\eta_i-\eta_j)^2+(\phi_i-\phi_j)^2}$.

\begin{figure}[tb]
\begin{center}
\epsfxsize= 3in  
\leavevmode
\epsfbox{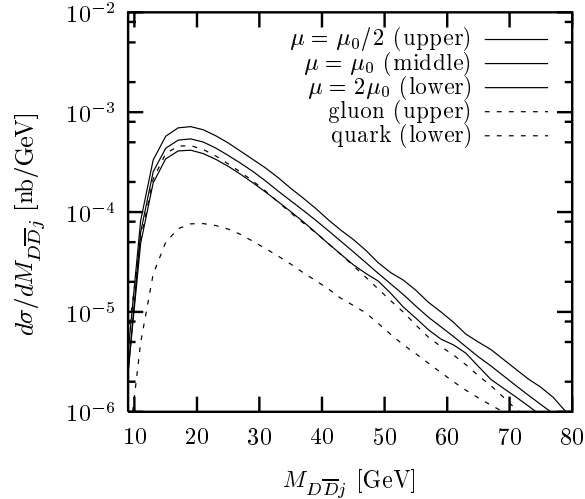}
\end{center}
\caption{Total cross section for the deep inelastic production of 
$D$--${\overline D}$--$jet$ as a function of their invariant mass 
$M_{D{\overline D}j}$ for $0<y<0.7$ and $2<Q^2<100\; {\rm GeV}^2$. 
The central scale is $\mu_0=\sqrt{Q^2+m_c^2+M^2_{D{\overline D}j}}$. 
A decomposition into quark and gluon initiated subprocesses is also shown.}
\label{fig3}
\end{figure}

\begin{figure}[tb]
\begin{center}
\epsfxsize= 4.5in  
\leavevmode
\epsfbox{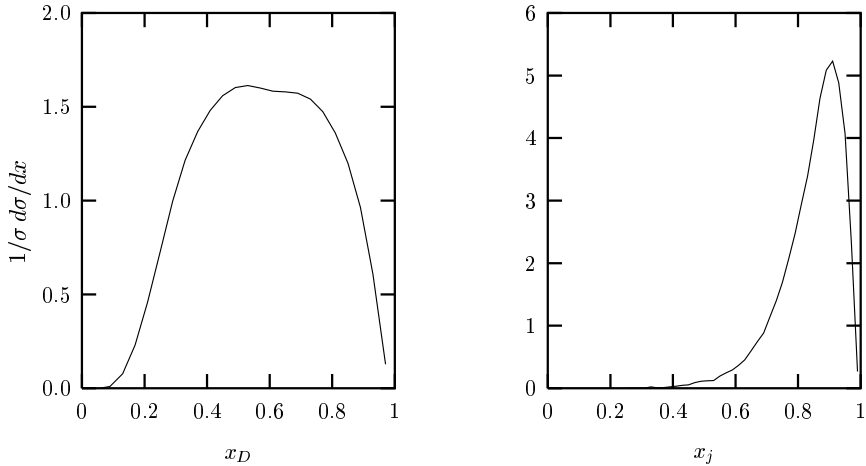}
\end{center}
\caption{Dependence of the $D{\overline D}j$ 
normalized cross section on the energy fraction $x$ 
for the $D$ hadron (left) and jet (right) for $0<y<0.7$ 
and $2<Q^2<100\; {\rm GeV}^2$.}
\label{fig4}
\end{figure}

The total cross section for the deep inelastic production of 
$D$--${\overline D}$--$jet$ as a function of their invariant mass 
$M_{D{\overline D}j}$ is shown in fig.~\ref{fig3} 
for $0<y<0.7$ and $2<Q^2<100\; {\rm GeV}^2$.
A one loop $\alpha_s$ with $n_f=3$ and $\Lambda_{QCD}^{(n_f=3)} = 232$ MeV, 
leading order GRV94\cite{grv94} parton distributions, and $m_c=1.35$ GeV 
where used. 
The normalization of this LO curve has a large uncertainty as 
demonstrated by the various scale choices $\mu=\{\mu_0/2,\mu_0,2\mu_0\}$, 
with $\mu_0=\sqrt{Q^2+m_c^2+M^2_{D{\overline D}j}}$.
Also shown in the figure is a decomposition into the gluon and light-quark 
initiated subprocesses.  The gluon initiated subprocess dominates 
due to the relatively large size of the gluon parton distribution function 
at small $x$.
As another example,
in the $D{\overline D}j$ center-of-mass frame we construct 
the Dalitz energy fractions $x_i=2E_i/M_{D{\overline D}j}$, 
($i=D,{\overline D}$, or $j$) that specify how much available energy 
is shared between the $D,{\overline D}$, and jet.  They satisfy 
$x_{D}+x_{\overline D}+x_j=2$.  The normalized cross section differential 
in $x_D$ and $x_j$ is shown in fig.~\ref{fig4}.

\section{Soft-gluon resummation}

As already remarked, existing NLO calculations for heavy quark 
electroproduction provide a solid theoretical perturbative QCD 
description \cite{lrsvn93,hs1,hs2}
for this reaction, and agree well with present data \cite{HERAc}.
At moderate $Q^2$ and $x$ values larger than $0.01$, the charm 
structure function $F_{2}^{\rm charm}$ is increasingly dominated by
partonic processes near the charm quark pair production threshold. 
The large size of the gluon density $f_g(x,\m)$ for small momentum 
fractions $x$ gives relatively large weight to such processes \cite{v96}. 
Although the QCD corrections at presently accessible $x$ values are 
moderate (about 30-40\%),
with an increasing number of data to be gathered at higher $x$,
it is worthwhile to have a closer look at such near-threshold subprocesses.
In this kinematic region, the QCD 
corrections are dominated by large Sudakov double logarithms.
Recently \cite{lm99}, these Sudakov logarithms have been resummed to 
all orders of 
perturbation theory, to next-to-leading logarithmic (NLL) accuracy, 
and, moreover, in single-particle inclusive (1PI) kinematics 
\cite{los98}\footnote{Analytical 
results for pair-inclusive kinematics are also given in \cite{lm99}}. 
Let us recall at this point the main results. First, the quality of 
the approximation for the {\it next-to-}leading logarithmic
threshold resummation was found to be clearly superior to leading 
logarithmic one.  Furthermore, the resummation provided NNLO estimates 
\cite{lm99},which were found to be sizable for $x\geq 0.05$. 

Below we give a synopsis of the analysis presented in \cite{lm99}.
We study electron proton scattering with the exchange 
of a single virtual photon, $Q^2=-q^2$, 
and a detected heavy quark (we concentrate on the charm quark
case here) in the final state, i.e. the subprocess
\beql
\g(q) +\, P(p) &\longrightarrow& {\rm{Q}}(p_1) +\, X[\bar{\rm{Q}}]\, ,
\label{elecprotscatt}
\eeql
where $X$ denotes any additional hadrons, including the heavy anti-quark,
in the final state and $p_1^2 =m^2$. 
The Mandelstam invariants, 
$S^{\prime} = (p + q)^2 + Q^2\,, T_1 = (p - p_1)^2-m^2 $ and  
$U_1 = (q -p_1)^2-m^2$ can be used to define $S_4=S^{\prime} + T_1 + U_1$,
which vanishes at the hadronic threshold.
The double differential heavy quark structure function $dF_2$ 
associated to (\ref{elecprotscatt}) may be written as
\beq
\frac{d^2F_{2,P}(x,S_4,T_1,U_1,Q^2,m^2)}{dT_1\,dU_1}
= {1\over S'^2}\sum_{i=q,\bar{q},g} \,\int\limits_{z^-}^{1}\frac{dz}{z}
\,f_{i/P}(z,\mu^2)\;
\omega_{2,i}\Big({x\over z},{s_4\over\mu^2},
                            {t_1\over\mu^2},
                            {u_1\over\mu^2},
                            {Q^2\over\mu^2},
                            {m^2\over\mu^2}, \alpha_s(\mu)
\Big) \;, 
\label{d2ffact}
\eeq
where $z^- = -U_1/(S'+T_1)$. The $f_{i/P}$ denote parton distributions in 
the proton at momentum fraction $z$ and $\Bar{{\rm{MS}}}$-mass 
factorization scale $\m$. The functions $\o_{2,i}$ describe the underlying hard parton 
scattering processes and depend on the partonic Mandelstam variables 
$s^{\prime}, t_1,u_1$ and $s_4$, which are derived from (\ref{elecprotscatt}) 
by replacing the proton $P$ by a parton of momentum $k= z p$. 
At $n$-th order in perturbation theory, the gluonic hard part $\o_{2,g}$ 
in eq.(\ref{d2ffact}) typically depends on singular distributions 
$\a_s^n [\ln^{2n-1-k}(s_4/m^2)/s_4]_+,\,k=0,1,\dots$, that must be resummed. 
Contributions from light initial-quark states are neglected, as they are
about 5\% at NLO.

The resummation of threshold logarithms rests upon 
the factorization of the kinematics and dynamics of the
relevant degrees of freedom near threshold \cite{s87,cls96}. 
The dynamical factors involved can be each be assigned a kinematic
weight $w_i$ that is defined to vanish at threshold.
For $d F_{2,P}$ in eq.(\ref{d2ffact}), the factorization of the
kinematics implies that these weights sum to the overall inelasticity
near threshold:
$S_4/m^2 \simeq w_1(-u_1)/m^2 + w_s$, with $w_1=1 - z$ and 
$w_s = s_4/m^2$. 
Correspondingly, the infrared regulated partonic structure 
function $d F_{2,g}$ factorizes into functions that individually organize 
contributions from these near-threshold degrees of freedom.
Thus, there is here a function $\j_{g/g}$ that sums the singular distributions from incoming
collinear gluons, and a soft function $S$ that organizes those
due to soft gluons not collinear to the incoming parton. Finally, 
there is a hard function $H_{2,g}$ incorporating only regular short-distances corrections.
Replacing the proton in eq.(\ref{d2ffact}) by a gluon, and passing to
Laplace-moment space, 
$\tilde{f}(N) = \int_0^{\infty}\! dw\, {\rm{exp}}[-N w] f(w)$, this gives \cite{lm99}
\beql
\tilde{\omega}_{2,g}
\left( N,{t_1\over \mu^2},{u_1\over \mu^2},{Q^2\over\mu^2},{m^2\over\mu^2}\right)
&=&
H_{2,g}\left({t_1\over \mu^2},{u_1\over \mu^2},{Q^2\over\mu^2},{m^2\over\mu^2}\right)\;
\left[ {{\tilde\psi}_{g/g}(N_u, p\cdot\zeta/\mu)
\over {\tilde\phi}_{g/g}(N_u,\mu)} \right] \;{\tilde S}\left({m\over N\mu},
\zeta\right)\; ,
\label{omegamom}
\eeql     
where $\f_{g/g}$ is the usual $\Bar{\rm MS}$-distribution
from mass factorization and $N_u=N(-u_1)/m^2$. 
In moment space, the Sudakov logarithms appear 
as factors $\a_s^n \ln^{2n-i}\!\!N$, with $i=0,1$ for NLL accuracy. The
$N$-dependence in eq.(\ref{omegamom}) exponentiates for each function 
individually. All leading logarithms (LL) are exclusively contained 
in ${\tilde{\j}}_{g/g}$, which is a gluon distribution at fixed fraction of $p\cdot \z$ and 
can be defined as an operator matrix element. 
It depends on a time-like vector $\z=p_2/m$ ($p_2$ is the heavy antiquark momentum).
Its collinear poles are canceled by $\f_{g/g}$. 
The threshold logarithms in ${\tilde{\j}}_{g/g}$ are resummed 
in analogy to the Drell-Yan process \cite{s87}, while all scale dependence 
of ${\tilde{\j}}_{g/g}$  and ${\tilde{\f}}_{g/g}$ is governed by 
renormalization group equations (RGE) 
with anomalous dimensions $\g_\j=2\g_g$ and $\g_{g/g}$ \cite{kos98,lm99}.

The soft function $S$ requires renormalization, 
since it is defined as a composite operator, that connects Wilson 
lines in the direction of the scattering partons \cite{kos98,ks97,kos982}. 
Its RGE, $\m (d/d\m)  \ln \tilde{S}(N) = - 2\, {\rm{Re}} \G_S$,
resums all threshold logarithms in $\tilde{S}$. Its gauge dependence cancels 
precisely that of $\j_{g/g}$. 
The soft anomalous dimension $\G_S$ is to order $\a_s$ 
\beql
\G_S(\a_s)&=&\frac{\a_s}{\p}
\Biggl\{ \!\left(\frac{C_A}{2} - C_F\right)\! ( L_\b + 1 ) - 
\frac{C_A}{2} \left( \ln \left({(p \cdot \z)^2\over m^2}\right) + 
\ln\frac{4\, m^4}{t_1\, u_1} \right)\! \Biggr\}\, ,
\label{softadim-res}
\eeql
with $\b = \sqrt{1- 4\, m^2/s}$ and 
$L_\b=(1-2\,m^2/ s)\{ \ln (1-\b)/(1+\b) + {\rm{i}}\p \}/\b$.

The final result for the hard scattering function  
$\tilde{\o}_{2,g}$ in moment space resums all large logarithms 
in single-particle inclusive kinematics up to NLL accuracy.
Combining the resummed $\tilde{\j}_{g/g}$ with the integrated 
RGE for $\tilde{S}$, 
we obtain for $\tilde{\o}_{2,g}$ \cite{lm99}
\beql
\label{sigNHSfinal}
\tilde{\omega}_{2,g}
\left( N,{t_1\over \mu^2},{u_1\over \mu^2},{Q^2\over\mu^2},{m^2\over\mu^2}\right)
&=& \;\\
&\ & \hspace{-50mm} \;
H_{2,g}\left({t_1\over m^2},{u_1\over m^2},{Q^2\over m^2},1\right)\;
{\tilde S}\left(1,\alpha_s({m\over N})\right)
\exp\Bigg \{ -2\int\limits_{\mu}^{m}{d\mu'\over\mu'}
\gamma_g\left(\alpha_s(\m^{\prime})\right)\Bigg\}
\nonumber \\
&\ & \hspace{-50mm} \times\;
\exp \Bigg \{ 
\int\limits_0^{\infty}\! \frac{d w}{w}  
\! \left(1 - {\rm{e}}^{-N_uw} \right)\! 
\Bigl[\, \int\limits_{w^2}^1 \frac{d \l}{\l} 
A_{(g)}(\a_s(\sqrt{\l} m)) + \frac{1}{2} \n_{(g)}(\a_s(w m)) \Bigr] 
\Bigg \}
\nonumber\\
&\ & \hspace{-50mm}\times\;
\exp\Bigg \{\int\limits_{m}^{m/N}{d\mu'\over\mu'}
2\, {\rm Re} \Gamma_S\left(\alpha_s(\m^{\prime})\right) \Bigg\}
\exp\Bigg \{ -2\int\limits_{\mu}^{2p\cdot \z}{d\mu'\over\mu'}
\left(\gamma_g\left(\alpha_s(\m^{\prime})\right)-
\gamma_{g/g}\Big( N_u,\alpha_s(\m^{\prime})\Big)\right)
\Bigg\}\, .\nonumber 
\eeql
The second exponent gives the leading $N$-dependence of the ratio 
${\tilde{\j}}_{g/g}/ {\tilde{\f}}_{g/g}$ 
with $\n_{(g)}(\a_s) = 2 C_A \a_s/\p$, 
$A_{(g)}(\a_s) = C_A (\a_s/\p) + (C_A K/2) (\a_s/\p)^2$  and 
$K=C_A(67/18-\p^2/6)-5/9n_f$ \cite{kt82}.
For NLL Sudakov resummation, the product $H_{2,g} \cdot S$ on the first
line of eq.~(\ref{sigNHSfinal}) is
determined from matching to the Born cross section at the scale $\m = m$. 

\begin{figure}[hbt]
\begin{center}
\epsfig{file=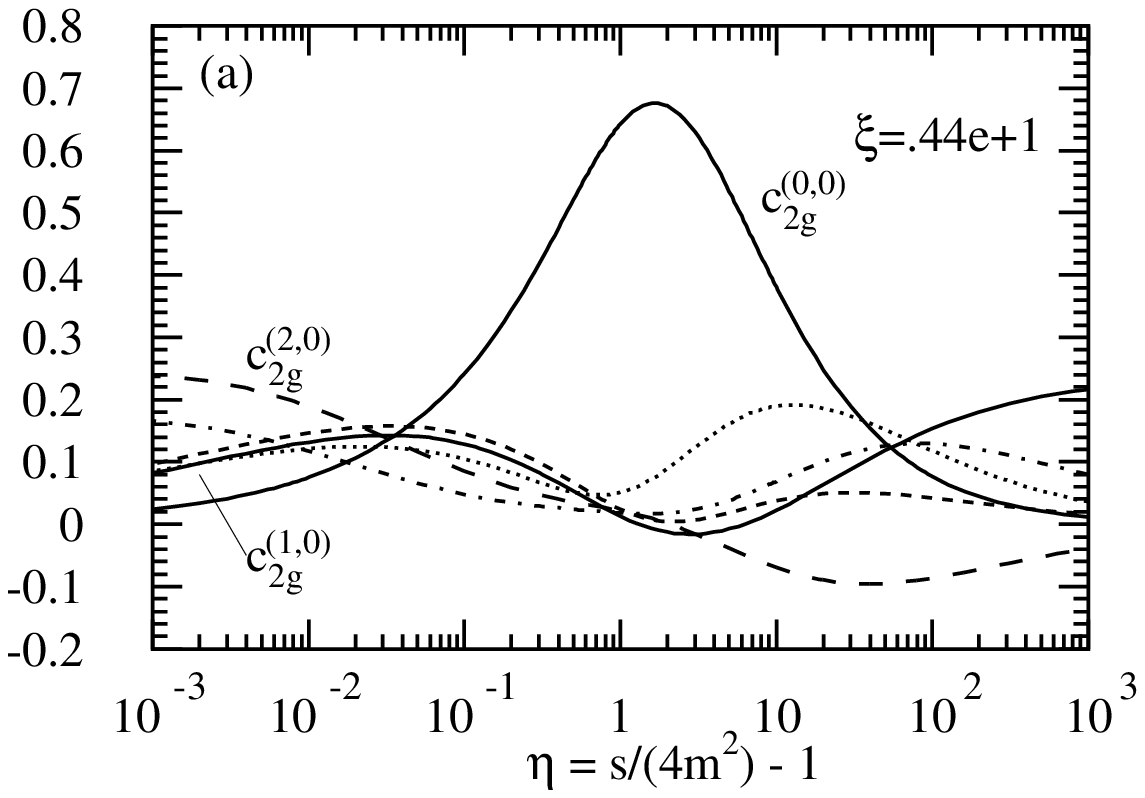,%
bbllx=50pt,bblly=130pt,bburx=285pt,bbury=470pt,angle=270,width=8.25cm}
\epsfig{file=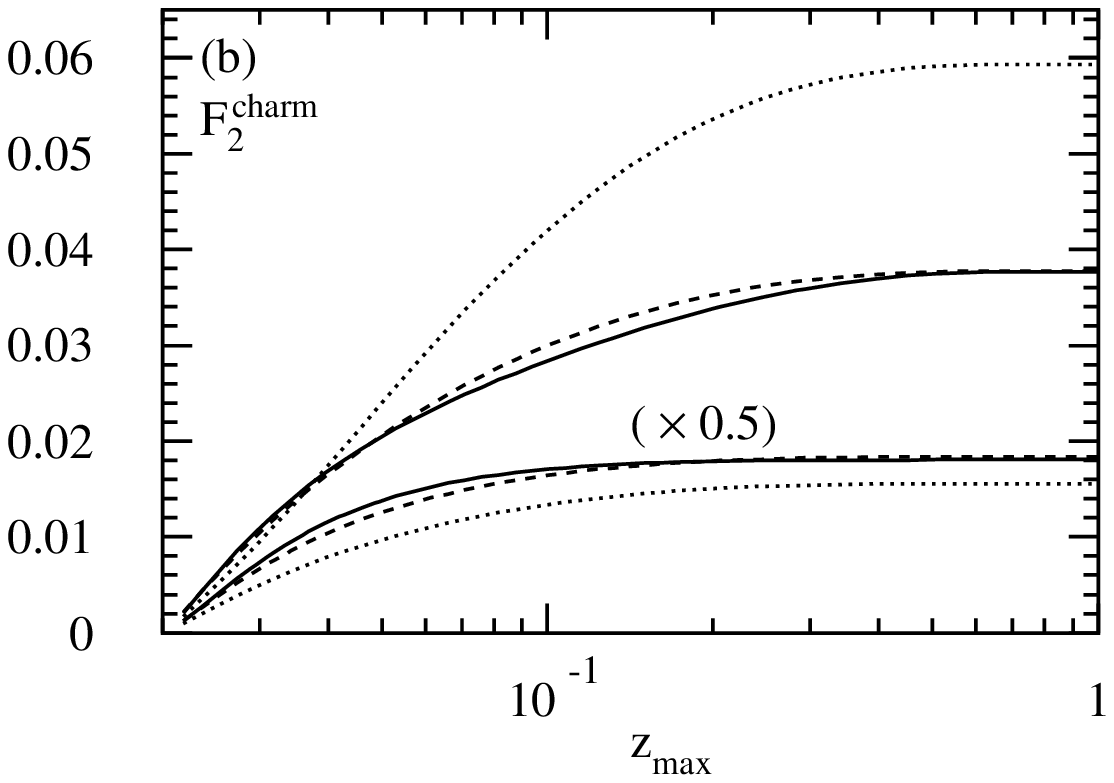,%
bbllx=50pt,bblly=110pt,bburx=285pt,bbury=450pt,angle=270,width=8.25cm}
\caption[dum]{\label{plot-one} {\small{
(a): The $\h$-dependence of the coefficient functions 
$c^{(k,0)}_{2,g}(\h,\x),\;k=0,1,2$   
for $Q^2=10\,{\rm GeV}^2$ with $m=1.5\,{\rm GeV}$. 
Plotted are the exact results for $c^{(k,0)}_{2,g},\;k=0,1$ 
(solid lines), the LL approximation to $c^{(1,0)}_{2,g}$ (dotted line) 
the NLL approximation to $c^{(1,0)}_{2,g}$ (dashed line), 
the  LL approximation to $c^{(2,0)}_{2,g}$ (dash-dotted line) and 
the NLL approximation to $c^{(2,0)}_{2,g}$ (long dashed line). 
(b): $F_2^{\rm charm}(x,Q^2,z_{\rm{max}})$ as a function of $z_{\rm{max}}$ 
at NLO  with the CTEQ4M gluon PDF, $x = 0.01$, 
$m =1.6\,{\rm GeV}$, $Q^2 = 10\,{\rm GeV}$ and $\m = m$ (upper three curves), 
$\m =\sqrt{Q^2 + 4 m^2}$ (lower three curves), rescaled by a factor of $0.5$.
Plotted are: The exact results (solid lines), 
the LL approximations (dotted lines) and the NLL approximations (dashed lines).}}}
\end{center}
\end{figure}

\begin{figure}[hbt]
\begin{center}
\epsfig{file=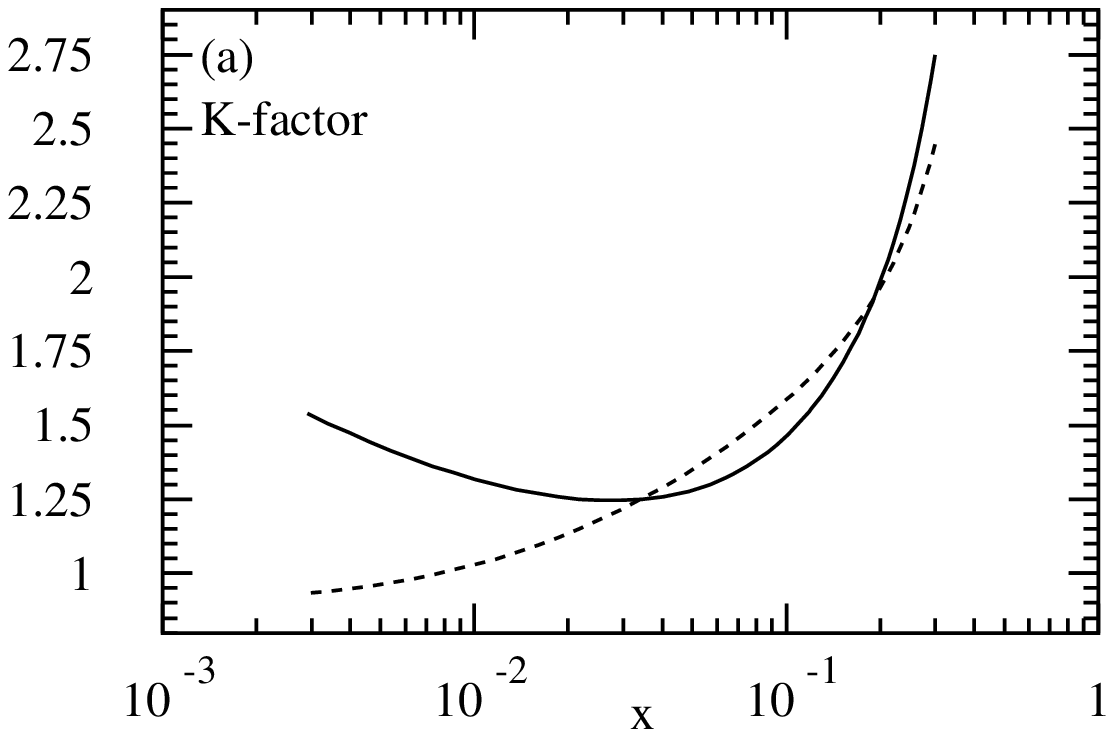,%
bbllx=50pt,bblly=110pt,bburx=285pt,bbury=450pt,angle=270,width=8.25cm}
\epsfig{file=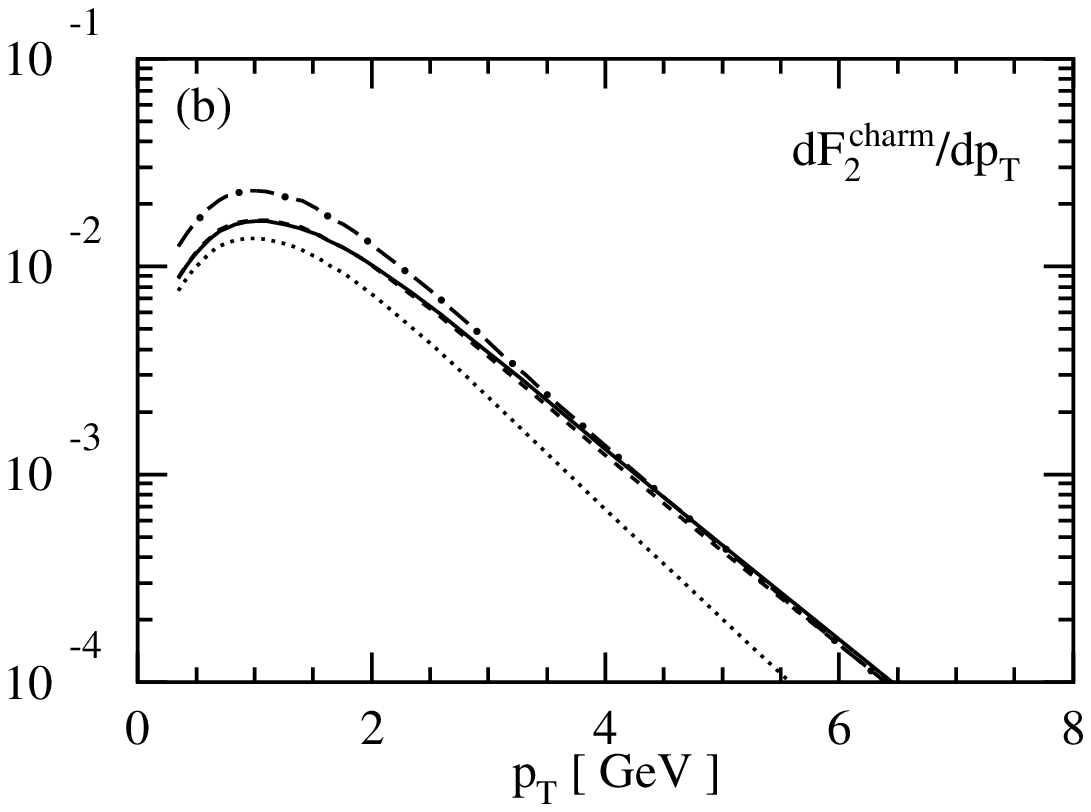,%
bbllx=50pt,bblly=130pt,bburx=285pt,bbury=470pt,angle=270,width=8.25cm}
\caption[dum]{\label{plot-two}{\small{
(a): The $x$-dependence of the ratios 
$F_{2\,\, (NLO)}^{\rm charm}/F_{2\,\, (LO)}^{\rm charm}$ (solid line) and 
$F_{2\,\, (NNLO)}^{\rm charm}/F_{2\,\, (NLO)}^{\rm charm}$ (dashed line) 
with 
$F_{2\,\, (NNLO)}^{\rm charm}$ in the improved NLL approximation (exact NLO result 
plus NLL approximate NNLO result) with CTEQ4M gluon PDF, 
$\m = m = 1.6\,{\rm GeV}$ and $Q^2 = 10\,{\rm GeV}$. 
(b): The differential distribution $d F_2^{\rm charm}/ d p_T$ 
as a function of $p_T$ 
with the CTEQ4M gluon PDF, $x = 0.01$, $m =1.6\,{\rm GeV}$, 
$Q^2 = 10\,{\rm GeV}$ and scale choice $\m = \sqrt{Q^2 + 4 (m^2 + p_T^2)}$.
Plotted are: The exact result (solid line) at NLO, 
the LL approximation at NLO (dotted line), the NLL approximation at NLO 
(dashed line) and at NNLO the improved NLL approximation (exact NLO result 
plus NLL approximate NNLO result) (long dash dotted line).
}}}
\end{center}
\end{figure}

The resummed result for $\tilde{\o}_{2,g}$ in eq.~(\ref{sigNHSfinal})
may be used as a generating functional for fixed order approximate perturbation 
theory by re-expanding $\tilde{\o}_{2,g}$ to NLO and NNLO 
and inverting the Laplace transform. 
After insertion of eq.~(\ref{sigNHSfinal}) into eq.~(\ref{d2ffact})
and integration over $T_1,U_1$, we may expand
the structure function as
\beql
F_2^{\rm charm}(x,Q^2,m^2) \!&=&\!
\frac{\a_s(\m)\, e_{{\rm c}}^2 Q^2}{4 \p^2 m^2}\! 
\int\limits_{ax}^{1}\, dz\, f_{g/P}(z,\mu^2)\,
\sum\limits_{k=0}^{\infty} (4 \p \a_s(\m))^k 
\sum\limits_{l=0}^{k} 
c^{(k,l)}_{2,g}(\h,\x) \ln^l\frac{\m^2}{m^2} , \,\,\,\,\,
\label{charmstrucintegrated}
\eeql
where $a=(Q^2+4m^2)/Q^2$ and $e_c=2/3$.

The quality of the NLL approximation eq.(\ref{sigNHSfinal}) can then 
be investigated by comparing  exact, LL and NLL approximation to 
the gluon coefficient functions $c^{(k,l)}_{2,g}$, which is done in 
fig.~\ref{plot-one}(a). 
The functions $c^{(k,l)}_{2,g}$ depend on the scaling variables
\beql
\h &=& \frac{s}{4 m^2}\, -1\, , \,\,\,\,\,\,  \,\,\,\,\,\, \,\,\,\,\,\,
\x \,=\, \frac{Q^2}{m^2}\, ,
\label{etaxidef}
\eeql
where $\h$ is a direct measure of the distance to the partonic threshold.  

Fig.~\ref{plot-one}(a) reveals that, although at one loop the LL accuracy provides
a good approximation for very small $\h$, the NLL approximation is excellent
over a much wider range in $\h$,
up to values of about 10 (the same actually holds true for
the $c^{(k,l)}_{2,g},\,l > 0, k\leq 2$).
We also show $c^{(2,0)}_{2,g}$, which 
has more structure than in the $c^{(1,0)}_{2,g}$ curves.

To address the threshold sensitivity of the integrated 
charm structure function 
to threshold processes, we perform the integral
over $z$ in eq.~(\ref{d2ffact}) only up to a value
$z_{\rm{max}}$, and plot the integral then as a function of $z_{\rm{max}}$. 
In this way we can see where the integral
eq.~(\ref{d2ffact}) acquires most of its value. 
The physical structure function corresponds
to ${z_{\rm{max}}}=1$.   
Fig.~\ref{plot-one}(b) demonstrates that for $x=0.01 $
$F_2^{\rm charm}$ is mostly determined by partonic processes 
close to threshold.        

In fig.~\ref{plot-two}a we display at a fixed value of the
factorization scale $\m = m$ over a range of
$x$, $0.003 \le x \le 0.3$ the effect of the
NNLO corrections. We plot the K-factors
$F_{2\,\, (NNLO)}^{\rm charm}/F_{2\,\, (NLO)}^{\rm charm}$ and,
for comparison, also  $F_{2\,\, (NLO)}^{\rm charm}/F_{2\,\, (LO)}^{\rm charm}$
\footnote{For $F_{2\,\, (LO)}^{\rm charm}$
we used a two-loop $\alpha_s$ and NLO gluon density.}.
At NNLO we have taken the improved NLL approximation to $F_2^{\rm charm}$
(the exact NLO result plus the NLL approximate NNLO result).
We see that particularly for smaller $x$, the size of the
NNLO corrections is negligible, the K-factor being close to one,
whereas for larger $x$, their overall size is still quite big,
almost a factor of 2 at $x=0.1$.
                                    
Finally, 
in fig.~\ref{plot-two}b we show the NLO results as a function of 
$p_T$ for $x=0.01$, $m =1.6\,{\rm GeV}$, $Q^2 = 10\,{\rm GeV}$ and 
$\m = \sqrt{Q^2 + 4 (m^2 + p_T^2)}$. At NLO, we compare our LL and NLL 
approximate results with the exact results of the second Ref.\ 
in \cite{lrsvn93}. 
We see that the exact curves are reproduced well
both in shape and magnitude by our NLL approximations,
whereas the curves for LL accuracy systematically underestimate
the true result. We also display the improved NLL approximation 
to $d F_2^{\rm charm}/ d p_T$ at NNLO, which contains sizeable contributions 
to the value of the maximum increases by $40\%$ - $50\%$.

\section{Conclusions}
Driven by the ever increasing variety and quantity of 
deep-inelastic charm production data from the H1 and ZEUS 
experiments at HERA, we have updated and reviewed two important 
tools: soft gluon threshold resummation and the next-to-leading 
order Monte-Carlo {\sc HVQDIS}.  The addition of semileptonic 
decay electron information and an option for only three body final states 
to the Monte-Carlo will enhance future physics analysis options.
Soft gluon threshold resummation, on the other hand, teaches us about the 
size of the terms neglected in fixed order calculations.

\subsection*{Acknowledgments}

The work of B.W.H. was supported by the U.~S.~Department of Energy, 
High Energy Physics Division, Contract No.\ W-31-109-Eng-38.  
The work of J.S. was supported in part by the U.~S. National Science 
Foundation grant PHY-9722101.
The work of S.M. and E.L. is part of the research program of the 
Foundation for Fundamental Research of Matter (FOM) and 
the National Organization for Scientific Research (NWO).

\end{document}